\documentclass[amsmath,prl,aps,twocolumn,superscriptaddress,floatfix,nofootinbib]{revtex4}

\usepackage{amsfonts}
\usepackage{bm}
\usepackage{graphicx}
\usepackage{color}

\newcommand{\be}{\begin{equation}}
\newcommand{\ee}{\end{equation}}
\newcommand{\ba}{\begin{eqnarray}}
\newcommand{\ea}{\end{eqnarray}}
\newcommand{\nn}{\nonumber}

\newcommand\lsim{\mathrel{\rlap{\lower4pt\hbox{\hskip1pt$\sim$}}
        \raise1pt\hbox{$<$}}}
\newcommand\gsim{\mathrel{\rlap{\lower4pt\hbox{\hskip1pt$\sim$}}
        \raise1pt\hbox{$>$}}}

\def\fnl{f_{NL}}

\def\taunl{\tau_{NL}}
\def\k{{\bf k}}
\def\q{{\bf q}}
\def\bigoh{{\mathcal O}}
\def\hP{{\hat P}}
\def\hA{{\hat A}}
\def\dtaunl{{\Delta\tau_{NL}}}
\def\hfnl{{\hat f_{NL}}}
\def\htaunl{{\hat\tau}_{NL}}
\def\ellmax{\ell_{\rm max}}

\begin{document}

\title{A universal bound on $N$-point correlations from inflation}

\author{Kendrick M. Smith}
\affiliation{Princeton University Observatory, Peyton Hall, Ivy Lane, Princeton, NJ 08544 USA}
\author{Marilena LoVerde}
\affiliation{Institute for Advanced Study, Einstein Drive, Princeton, NJ 08540, USA}
\author{Matias Zaldarriaga}
\affiliation{Institute for Advanced Study, Einstein Drive, Princeton, NJ 08540, USA}

\date{\today}


\begin{abstract}
Models of inflation in which non-Gaussianity is generated outside the horizon, such as curvaton models,
generate distinctive higher-order correlation functions in the CMB and other cosmological observables.
Testing for violation of the Suyama-Yamaguchi inequality $\taunl \ge (\frac{6}{5} \fnl)^2$, where $\fnl$ and $\taunl$ denote 
the amplitude of the three-point and four-point functions in certain limits, has been proposed as a way to distinguish
qualitative classes of models.
This inequality has been proved for a wide range of models, but only weaker versions have been proved in general.
In this paper, we give a proof that the Suyama-Yamaguchi inequality is always satisfied.
We discuss scenarios in which the inequality may appear to be violated in an experiment such as Planck, and how
this apparent violation should be interpreted.
We analyze a specific example, the ``ungaussiton'' model, in which leading-order scaling relations
suggest that the Suyama-Yamaguchi inequality is eventually violated, and show that the inequality always holds.
\end{abstract}

\maketitle


A central goal of observational cosmology is to characterize the statistics of the initial
perturbations of our universe, thereby constraining the physics
which generated these perturbations.
Observations to date are consistent with adiabatic, Gaussian, scalar initial conditions, but
deviation from a scale invariant initial power spectrum has been observed with 3$\sigma$ significance
\cite{Komatsu:2010fb}.

The search for non-Gaussian statistics in the initial fluctuations has emerged as a particularly
interesting probe of inflation, due to the presence of many distinct signals (or ``shapes'') which
probe inflationary physics in different ways.
For example, detection of a nonzero three-point correlation function $\langle \zeta_{\k_1} \zeta_{\k_2} \zeta_{\k_3} \rangle$
in ``squeezed'' configurations ($k_1 \ll \min(k_2,k_3)$) would rule out all single-field models of inflation
\cite{Creminelli:2004yq}, while detection of an
``equilateral'' three-point function ($k_1\sim k_2\sim k_3$) is a generic test for self-interactions of the inflaton \cite{Senatore:2009gt}.

Throughout this paper, we use the following notation.
We denote the connected part of an $N$-point expectation value
by $\langle \cdot \rangle_c$, and use the notation $\langle\cdot\rangle'$ 
to denote the expectation value without the multiplicative factor $(2\pi)^3 \delta^3(\sum\k_i)$.
For a 3D field $\sigma$, we write $\Delta_\sigma^2(k) = k^3 P_\sigma(k)/2\pi^2$, 
where $P_\sigma(k)$ is the power spectrum.

The ``local'' model is a simple non-Gaussian model in which the initial adiabatic curvature is given by
$\zeta = \zeta_G + \frac{3}{5} \fnl (\zeta_G^2 - \langle\zeta_G^2\rangle)$, where $\zeta_G$ is a
Gaussian field and $\fnl$ is a free parameter.  The three-point and four-point functions in this model are:
\ba
\langle \zeta_{\k_1} \zeta_{\k_2} \zeta_{\k_3} \rangle' &=& \frac{6}{5} \fnl ( P_{k_1} P_{k_2} + \mbox{cyc.} ) \label{eq:taunl_shape} \\
\langle \zeta_{\k_1} \zeta_{\k_2} \zeta_{\k_3} \zeta_{\k_4} \rangle'_c &=& \taunl ( P_{k_1} P_{k_3} P_{|\k_1+\k_2|} + \mbox{11 perm.} ) \nn
\ea
where $P_k = P_\zeta(k)$ and $\taunl = ( \frac{6}{5} \fnl )^2$.

The local model can be generalized by introducing $N$ Gaussian fields $\sigma_1,\ldots,\sigma_N$
(assumed for simplicity to be uncorrelated with equal power spectra)
and taking the initial adiabatic curvature to be 
$\zeta = A_i \sigma_i + B_{jk} (\sigma_j \sigma_k - \langle \sigma_j \sigma_k \rangle)$.
This also gives rise to non-Gaussianity of local type~(\ref{eq:taunl_shape}), but the
relation between three-point and four-point functions is relaxed to an inequality:
\be
\taunl \ge \left( \frac{6}{5} \fnl \right)^2
\ee
Suyama and Yamaguchi showed \cite{Suyama:2007bg} that this inequality is always true
at tree level in the $(\Delta N)$ expansion.
In~\cite{Sugiyama:2011jt}, it was observed that individual loop diagrams can violate
the inequality, raising the interesting question of whether violation of the Suyama-Yamaguchi
inequality could be an observational signature of loop diagrams in inflation.

In this paper, we will answer this question negatively: we will give a proof that the SY
inequality is always satisfied.  The proof is very general and does not depend on any physics;
it comes from the requirement that a suitably constructed covariance matrix is always positive
definite.  We will also discuss scenarios in which the inequality can appear to be violated
observationally, and how this should be interpreted.  

\section{General proof that $\taunl \ge (6\fnl/5)^2$}

The intuitive idea behind this proof is that if we define a local estimate of the small-scale
power $\hP$, then $\taunl$ is the auto power of $\hP$ on large scales, whereas $(6\fnl/5)^2$
is the part of the auto power which can be attributed to the cross-correlation of $\hP$ with $\zeta$.

To give the proof in maximum generality, we generalize the definitions of $\fnl$ and $\taunl$
by taking the squeezed limit ($k_1 \ll k_2$) of the three-point function and the collapsed limit
($|\k_1+\k_2| \ll \min(k_i)$) of the four-point function:
\ba
\fnl &=& \frac{5}{12} \lim_{k_1\rightarrow 0} \frac{\langle \zeta_{\k_1} \zeta_{\k_2} \zeta_{\k_3} \rangle'}{P_\zeta(k_1) P_\zeta(k_2)}  \label{eq:fnl_def}  \\
\taunl &=& \frac{1}{4} \lim_{|\k_1+\k_2|\rightarrow 0} 
  \frac{\langle \zeta_{\k_1} \zeta_{\k_2} \zeta_{\k_3} \zeta_{\k_4} \rangle'_c}{P_\zeta(k_1) P_\zeta(k_3) P_\zeta(|\k_1+\k_2|)}  \label{eq:taunl_def}
\ea
This defines $\fnl$ and $\taunl$ for arbitrary non-Gaussian initial conditions.
In the special case where the three-point and four-point functions have local shapes~(\ref{eq:taunl_shape}), we recover
the usual definitions of $\fnl$ and $\taunl$.

For wavenumbers $k_L \ll k_S$, let $b_S$ be a narrow band of wavenumbers near $k_S$.
Define a field $\hP_{\k}$ by:
\be
\hP_{\k} = \frac{1}{V_S} \int_{\k'\in b_S} \frac{d^3\k'}{(2\pi)^3} \frac{\zeta_{\k'} \zeta_{\k-\k'}}{P(k')}
\ee
where $V_S = \int_{\k'\in b_S} d^3\k'/(2\pi)^3$ is the volume of the band. The field $\hP_{\k}$
represents the long-wavelength variation in the locally measured small-scale power. 

The power spectrum of $\hP_{\k}$ and its cross power spectrum with $\zeta_{\k}$ are given by:
\ba
\langle \hP_{\k_L}^* \hP_{\k_L} \rangle' &=&
\frac{1}{V_S^2} \int \frac{d^3\k' d^3\k''}{(2\pi)^6} \frac{\langle \zeta_{-\k'} \zeta_{\k'-\k_L} \zeta_{\k''} \zeta_{-\k''+\k_L} \rangle_c'}{P_\zeta(k') P_\zeta(k'')} \nn \\
  && \hspace{0.5cm} + \frac{1}{V_S^2} \int \frac{d^3\k'}{(2\pi)^3} \frac{2 P_\zeta(k') P_\zeta(k'')}{P_\zeta(k') P_\zeta(k'')} \nn \\
& \rightarrow & 4 \taunl P_\zeta(k_L) + \frac{2}{V_S}  \label{eq:PP_squeezed} \\
\langle \zeta_{\k_L}^* \hP_{\k_L} \rangle'
  &=& \frac{1}{V_S} \int \frac{d^3\k'}{(2\pi)^3} \frac{\langle \zeta_{-\k_L} \zeta_{\k'} \zeta_{\k_L-\k'} \rangle}{P_\zeta(k')} \nn \\
  & \rightarrow & \frac{12}{5} \fnl P_\zeta(k_L)  \label{eq:zP_squeezed}
\ea
where ``$\rightarrow$'' denotes the leading behavior of each term in the $k_L \ll k_S$ limit.

Now consider the covariance matrix of the fields $\zeta,\hP$:
\be
\left( \begin{array}{cc}
  \langle \zeta_{\k_L}^* \zeta_{\k_L} \rangle' & \langle \zeta_{\k_L}^* \hP_{\k_L} \rangle' \\
  \langle \zeta_{\k_L} \hP_{\k_L}^* \rangle' & \langle \hP_{\k_L}^* \hP_{\k_L} \rangle'
\end{array} \right)
\ee
The determinant must be positive.  Plugging in
Eqs.~(\ref{eq:PP_squeezed}),~(\ref{eq:zP_squeezed}), we get:
\be
\taunl \ge \left( \frac{6}{5} \fnl \right)^2 - \frac{1}{2 P_\zeta(k_L) V_S}   \label{eq:sy_fv}
\ee
Now take the limit $(k_L/k_S) \rightarrow 0$.  The second term on the RHS (which represents the disconnected part
of the four-point function in Eq.~(\ref{eq:PP_squeezed})) goes to zero, and we obtain the SY inequality
\be
\taunl \ge \left( \frac{6}{5} \fnl \right)^2  \label{eq:qed}
\ee
This completes the proof.
The proof is valid for initial conditions with an arbitrary non-Gaussian PDF, and makes 
no assumptions about the physics.
The SY inequality emerges as a positivity constraint which is always satisfied, in
the same sense that the power spectrum of any field is constrained to be nonnegative.
The SY inequality was also interpreted recently as a positivity constraint in \cite{Lewis:2011au}.

The proof above assumes that $\fnl$ and $\taunl$ are defined in squeezed limits.
If we define them at fixed scales $k_L, k_S$, then Eq.~(\ref{eq:sy_fv}) shows that the SY inequality can be
violated by an amount $\Delta\taunl = 1/(2P_\zeta(k_L) V_S)$.  If we assume that $V_S = \bigoh(k_S^3)$,
then this means that subleading contributions to the four-point function can violate the SY inequality, but such
contributions must scale as $\Delta\taunl = \bigoh(k_L^3/k_S^3)$ in the collapsed limit.
Given our definition of $\taunl$, this corresponds to a four-point function
$\langle \zeta_{\k_1} \zeta_{\k_2} \zeta_{\k_3} \zeta_{\k_4} \rangle$ 
which is finite in the collapsed limit.
Conversely, a four-point function which is finite in the collapsed limit can generally have either sign
and violate the SY inequality.
Such examples can be interpreted as ``accidental'' $\taunl$ contributions from a four-point shape which is
very different from the $\taunl$ shape.  The amount by which the SY inequality can be violated
is very small and difficult to detect with statistical significance in an experiment such as Planck, 
as we discuss next.


\section{Estimators}

We now study the question: if we evaluate estimators for $\fnl$ and $\taunl$ in an experiment,
do there exist realizations of the data which appear to violate the inequality $\taunl \ge (\frac{6}{5} \fnl)^2$, 
and if so, how should we interpret such a measurement?

First consider an ideal experiment in which all modes $\zeta_{\k}$ are measured without noise, over some range of
scales in a periodic box.
Given this data, we can write estimators for $\fnl$ and $\taunl$ as follows.
We normalize Fourier transforms so that $\langle \zeta_{\k}^* \zeta_{\k'} \rangle = P_\zeta(k) V \delta_{\k\k'}$,
where $V$ is the box volume.
Define a field $\hP_{\k}$ by:
\be
\hat P_{\k} = \frac{1}{N_S} \sum_{\k' \in b_S} \frac{\zeta_{\k'} \zeta_{\k-\k'}}{P_\zeta(k')}
\ee
and estimators for $\fnl$ and $\taunl$ by:
\ba
\hfnl &=& \frac{5}{12 V N_L} \sum_{\k\in b_L} \frac{\zeta_{\k}^* \hat P_{\k}}{P_\zeta(k)} \\
\htaunl &=& \frac{1}{4 V N_L} \sum_{\k\in b_L} \left( \frac{\hat P_{\k}^* \hat P_{\k}}{P_\zeta(k)} - \frac{2 V^2}{P_\zeta(k) N_S} \right)  \label{eq:htaunl_def}
\ea
Here, $b_L,b_S$ are bins of wavenumbers near characteristic scales $k_L\ll k_S$, and $N_\alpha = \sum_{\k\in\alpha}1$ is the
number of modes in the bin (where $\alpha\in\{L,S\}$).
These definitions are closely analogous to the ones from the previous section, but here we are defining estimators which are
applied to a single realization $\zeta_{\k}$, rather than taking an ensemble average over realizations.

Since the sum of positive definite matrices is positive definite,
we have the positivity constraint:
\be
\mbox{Det} \sum_{\k} \frac{1}{P_\zeta(k)} \left( \begin{array}{cc}
  \zeta_{\k}^* \zeta_{\k} & \zeta_{\k}^* \hP_{\k} \\
  \zeta_{\k} \hP_{\k}^* & \hP_{\k}^* \hP_{\k}
\end{array} \right) \ge 0
\ee
which gives:
\be
\htaunl \ge \left( \frac{6}{5} \frac{\hfnl}{\hA} \right)^2 - \dtaunl  \label{eq:sy_finite}
\ee
where we have defined
\ba
\hA &=& \left( \frac{1}{V N_L} \sum_{\k\in b_L} \frac{\zeta_{\k}^* \zeta_{\k}}{P_\zeta(k)} \right)^{1/2} \\
\dtaunl &=& \frac{V}{2 N_L N_S} \sum_{\k\in b_L} \frac{1}{P_\zeta(k)}  \label{eq:dtaunl_def}
\ea
Eq.~(\ref{eq:sy_finite}) is the ``estimator'' version of the SY inequality.
It applies to each realization $\zeta_{\k}$ individually
(i.e.~one does not need to average over an ensemble of realizations for the inequality
to apply).

The quantity $\hA$ which appears can be interpreted as
a modification of the $\fnl$ estimator $(\hfnl \rightarrow \hfnl/\hA)$ which is necessary for
an estimator inequality to apply.
Note that a similar modification was proposed in \cite{Creminelli:2006gc}
as a way to reduce the estimator variance in the case where $\fnl$ is detected with statistical significance.

The quantity $\dtaunl$ is the maximum amount by which the estimator inequality $\htaunl \ge (\frac{6}{5} \hfnl/\hA)^2$
can be violated by an individual realization.
For the ideal experiment, $\dtaunl$ is equal to the quantity $1/(2 P_\zeta(k_L) V_S)$ obtained previously in 
Eq.~(\ref{eq:sy_fv}).

It is interesting to compare $\dtaunl$ to the statistical error $\sigma(\taunl)$ which can be obtained
using the estimator in Eq.~(\ref{eq:htaunl_def}).
A short calculation shows:\footnote{This calculation
makes the approximation that the connected four-point function of the field $\hP_{\k}$ is zero on large scales.
This type of approximation is common when forecasting higher-point estimators, e.g.~in the context of CMB lens
reconstruction, an analogous approximation has been shown to be accurate in \cite{Hanson:2010rp}.}
\be
\sigma(\taunl)^2 = \frac{V^2}{2 N_L^2 N_S^2} \sum_{\k\in b} \frac{1}{P_\zeta(k)^2}
\ee
The ratio $\dtaunl/\sigma(\taunl)$ is roughly $N_L^{1/2}$.  Assuming that the number of modes $N_L$ is $\gtrsim 10$,
we conclude that {\em in an ideal experiment, it is possible for a specific realization $\zeta_{\k}$ to violate the inequality 
$\taunl \ge (\frac{6}{5} \fnl)^2$ with statistical significance}.

How would we interpret such a violation if observed?  
As previously remarked, if we measure $\fnl$ and $\taunl$ at fixed
scales $k_L,k_S$, the SY inequality can be violated by subleading
terms which scale as $\taunl \sim \bigoh(k_L^3/k_S^3)$.
This scaling corresponds to a four-point function which is finite
in the collapsed limit.
What we have shown here is that an ideal experiment can ``see'' such
terms with statistical significance.
In this scenario, the natural interpretation is that the $\taunl$ estimator is receiving
accidental contributions from a four-point signal which is not the $\taunl$ shape.

The preceding discussion has considered an ideal experiment; let us now consider a more realistic case.
For concreteness, consider a cosmic variance limited CMB experiment with $\ellmax=400$.
The statistical error $\sigma(\taunl)_{\rm CMB}$ for this experiment is $\approx 4000$.
There will also be an estimator inequality of the form $\htaunl \ge (\frac{6}{5} \hfnl/\hA)^2 - (\dtaunl)_{\rm CMB}$.
Through numerical experiments (using a gradient minimization procedure to search for a realization which
violates the SY inequality as much as possible), we find that $(\dtaunl)_{\rm CMB} \approx 40000$ for this experiment.
Since $(\dtaunl)_{\rm CMB} \gsim \sigma(\taunl)_{\rm CMB}$, it is possible to find CMB realizations in which the
SY inequality appears to be violated with statistical significance.
For example, we can find realizations for which $\taunl = -40000 \pm 4000$, i.e.~$\taunl$ is negative at 
10$\sigma$.
Note that the ratio $(\dtaunl)/\sigma(\taunl)$ is roughly $N_L^{1/2}$, where $N_L$ is
the number of ``squeezed'' CMB multipoles with significant signal-to-noise; this number was found to be $N_L\approx 100$
in \cite{Kogo:2006kh}.

When we apply a CMB estimator $\htaunl$ to this dataset, we are measuring $\taunl$
averaged over a range of scales $k_L,k_S$, with a weighting which depends on CMB transfer
functions (as opposed to our previously assumed flat weighting) but is peaked roughly at $k_L\sim 0.0004$ Mpc$^{-1}$
and $k_S\sim 0.02$ Mpc$^{-1}$.
Now let us imagine we could do an ideal experiment in which we estimate $\taunl$ using all 3D modes $\zeta_{\k}$ throughout our Hubble
volume, {\em with the same weighting in $k_L,k_S$}.
In this ideal experiment, there will be a postivity constraint $\htaunl \ge (\frac{6}{5} \hfnl/\hA)^2 - (\dtaunl)_{\rm ideal}$,
where $(\dtaunl)_{\rm ideal} \approx (k_L/k_S) (\dtaunl)_{\rm CMB} = 800$.\footnote{To see that 
$(\dtaunl)_{\rm ideal} \approx (k_L/k_S) (\dtaunl)_{\rm CMB}$, we argue as follows.  For narrow bins
in $k_L,k_S$, Eq.~(\ref{eq:dtaunl_def}) gives $(\dtaunl)_{\rm ideal} = (k_L^3/2\Delta_{\zeta}^2 k_S^3) (\Delta\log k_S)$.
If we approximate the cosmic variance limited CMB as an ideal 2D measurement, then the value of $\dtaunl$ will
just be the 2D version of this, i.e.~$(\dtaunl)_{\rm CMB} = (k_L^2/2\Delta_{\zeta}^2 k_S^2) (\Delta\log k_S)$.
This shows that $(\dtaunl)_{\rm ideal} \approx (k_L/k_S) (\dtaunl)_{\rm CMB}$ for narrow bins; the general case
follows by integrating over $k_L,k_S$ with the appropriate weighting.}
Since this constraint is satisfied for every realization $\zeta_{\k}$, we do not actually need to do the ideal experiment in order to
conclude that the inequality $\taunl \gsim -800$ applies!
Although there exist CMB realizations with $\taunl=-40000$, it is mathematically impossible to extend such a realization from the
surface of last scattering to the Hubble volume; the extension automatically satisfies $\taunl \gsim -800$.

Given this picture, the only way to reconcile a measurement $\taunl^{\rm CMB} = -40000 \pm 4000$
(or more generally, a CMB measurement which violates $\taunl \ge (\frac{6}{5} \fnl)^2$ with statistical
significance)
with positivity constraints
seems to be to relax the assumption of translation invariance.
More precisely, translation invariance must be broken in a specific way where the
four-point function near our surface of last scattering is very different from its mean value throughout
the Hubble volume (so that $\taunl$ can be $\approx-40000$ near the surface of last scattering, and
have mean value $\gsim -800$ in the Hubble volume).
Since translation invariance is assumed in the construction of $\taunl$ estimator,
an observed violation of the SY inequality in the CMB would not be evidence for a particular model of inflation,
but rather a sign that something is wrong with the assumptions which motivated searching for $\taunl$ in the first place.
In this sense, testing the inequality with the CMB is more of a sanity check on the whole inflationary framework than a discriminator between models.

The preceding analysis applies to any experiment where $(\dtaunl) \gsim \sigma(\taunl) \gsim (\dtaunl)_{\rm ideal}$.
For example, this is the case for the upcoming Planck mission, since $\dtaunl/\sigma(\taunl)\approx 10$ and
$(\dtaunl) / (\dtaunl)_{\rm ideal} \approx k_L/k_S \approx 300$.
In this case, it is possible to find CMB realizations in which the estimators violate the inequality 
$\htaunl \ge (\frac{6}{5}\hfnl/\hA)^2$ with statistical significance, but it is mathematically inconsistent to interpret
this as evidence for violation of the inequality $\taunl \ge (\frac{6}{5}\fnl)^2$ throughout the Hubble
volume.

\section{Ungaussiton model}

The so-called ``ungaussiton'' model from \cite{Suyama:2008nt} is a multifield model of inflation in which the three-point and
four-point functions satisfy the following scaling relation in the limit of weak non-Gaussianity:
\be
\taunl = C \Delta_\zeta^{-2/3} \fnl^{4/3}  \label{eq:ug_scaling}
\ee
where $C$ is a constant of order 1. 
Extrapolating to large $\fnl$, this scaling relation suggests that the inequality $\taunl \ge (\frac{6}{5}\fnl)^2$ is violated
for $\fnl \gtrsim \Delta_\zeta^{-1}$.

As a check on our general theorem, in this section we will show explicitly that the Suyama-Yamaguchi inequality is always satisfied
in this model.  We will find that the scaling relation~(\ref{eq:ug_scaling}) is only valid for $\fnl \ll \Delta_\zeta^{-1}$,
and takes a different form (which always satisfies the SY inequality) for larger values of $\fnl$.
Calculation of the three-point and four-point functions in the ungaussiton model also appeared recently in \cite{Suyama:2010uj}.

In the ungaussiton model, the initial adiabatic curvature is of the form
\be
\zeta = N_\phi \phi + \frac{1}{2} N_{\sigma\sigma} (\sigma^2 - \langle\sigma^2\rangle)
\ee
where $N_\phi,N_{\sigma\sigma}$ are free parameters and the fields $\phi,\sigma$ are uncorrelated Gaussian fields with power spectra
given by $(k^3/2\pi^2) P_\sigma(k) = (H_I/2\pi)^2$, where $H_I$ is the Hubble constant during inflation. 
This is a slight simplification of the scenario considered
in \cite{Suyama:2008nt} but contains all the qualitative features, including the $\taunl = \bigoh(\fnl^{4/3})$ scaling at leading order in $\fnl$.

The power spectrum in this model is:
\be
P_\zeta(k) = N_\phi^2 P_\sigma(k) + \frac{N_{\sigma\sigma}^2}{4} P_{\sigma^2}(k)
\ee
where
\be
P_{\sigma^2}(k) = 2 \int \frac{d^3\q}{(2\pi)^3} P_\sigma(q) P_\sigma(|\k-\q|)
\ee
As a technical point, we note that if $P_\sigma(k)$ is scale-invariant, then the power spectrum $P_{\sigma^2}(k)$ 
is infrared divergent.  If the IR divergence is regulated by putting the fields in a finite box with length $L$, then the power spectrum
diverges as $P_{\sigma^2}(k) \sim 8\pi^2 \log(kL) \Delta_{\sigma}^4 / k^3$.
In what follows, we have left the regulator implicit, by writing all IR-divergent quantities in terms of 
$P_{\sigma^2}(k)$.

A short calculation shows that the squeezed three-point function and collapsed four-point function are given by:
\ba
\langle \zeta_{\k_1} \zeta_{\k_2} \zeta_{\k_3} \rangle' & \rightarrow & \frac{N_{\sigma\sigma}^3}{2} P_{\sigma^2}(k_1) P_\sigma(k_2) \\
\langle \zeta_{\k_1} \zeta_{\k_2} \zeta_{\k_3} \zeta_{\k_4} \rangle'_c & \rightarrow & N_{\sigma\sigma}^4 P_{\sigma^2}(|\k_1+\k_2|) P_\sigma(k_1) P_\sigma(k_3) \nn
\ea
Plugging the above expressions into the definitions~(\ref{eq:fnl_def}),~(\ref{eq:taunl_def}) of $\fnl$ and $\taunl$, we find:
\ba
\frac{6}{5} \fnl &=& \frac{\alpha^3 \beta(k_L)}{(1 + \alpha^2 \beta(k_L))(1 + \alpha^2 \beta(k_s))} \label{eq:ug_fnl} \\
\taunl &=& \frac{\alpha^4 \beta(k_L)}{(1 + \alpha^2 \beta(k_L))(1 + \alpha^2 \beta(k_s))^2} \label{eq:ug_taunl}
\ea
where we have defined dimensionless parameters $\alpha = N_{\sigma\sigma}/N_\phi^2$ and $\beta(k) = N_\phi^2 P_{\sigma^2}(k) / (4 P_\sigma(k))$.

Note that we recover the scaling relation~(\ref{eq:ug_scaling}) in the Gaussian limit
(i.e.~$\alpha^2\beta \ll 1$ with $\beta=\bigoh(\Delta_\zeta^2)$).
However, in the general form in Eqs.~(\ref{eq:ug_fnl}),~(\ref{eq:ug_taunl}), it may be seen that the SY inequality is always satisfied.
Indeed,
\be
\frac{\taunl}{(6\fnl/5)^2} = 1 + \frac{1}{\alpha^2 \beta(k_L)}
\ee
and the RHS is always $\ge 1$ since $\beta(k_L)$ must be positive.


\vspace{0.5cm}

{\em Acknowledgements.}
We would like to thank Eiichiro Komatsu and David Spergel for useful discussions.
KMS is supported by a Lyman Spitzer fellowship in the Department of Astrophysical Sciences at Princeton University.
ML is supported as a Friends of the Institute for Advanced Study Member and by the NSF though AST-0807444.
MZ is supported by the NSF under PHY-0855425, AST-0506556 and AST-0907969, by the David and Lucile Packard Foundation
and by the John D.~and Catherine T.~MacArthur Foundation.

\bibliographystyle{prsty}
\bibliography{ngac}

\end{document}